# Colloidal quasi-2D Methylammonium Lead Bromide Perovskite Nanostructures with Tunable Shape and High Chemical Stability

Eugen Klein,[1] Rostyslav Lesyuk,[1,2] Christian Klinke[1,3,4]*

[1] *Institute of Physics, University of Rostock, Albert-Einstein-Straße 23, 18059 Rostock, Germany*
[2] *Pidstryhach Institute for applied problems of mechanics and mathematics of NAS of Ukraine, Naukowa str. 3b, 79060 Lviv, Ukraine*
[3] *Department "Life, Light & Matter", University of Rostock, Albert-Einstein-Straße 25, 18059 Rostock, Germany*
[4] *Department of Chemistry, Swansea University – Singleton Park, Swansea SA2 8PP, United Kingdom*
* Corresponding author: christian.klinke@uni-rostock.de

Dedicated to Prof. Alexander Eychmüller of the TU Dresden on the occasion of his retirement.

**ABSTRACT**

Control over the lateral dimensions of colloidal nanostructures is a complex task which requires a deep understanding of the formation mechanism and reactivity in the corresponding systems. As a result, it provides a well-founded insight to the physical and chemical properties of these materials. In this work, we demonstrate the preparation of quasi-2D methylammonium lead bromide nanostripes and discuss the influence of some specific parameters on the morphology and stability of this material. The variation in the amount of the main ligand dodecylamine gives a large range of structures beginning with 3D brick-like particles at low concentrations, nanostripes at elevated and ultimately nanosheets at large concentrations. The amount of the co-ligand trioctylphosphine can alter the width of the nanostripe shape to a certain degree. The thickness can be adjusted by the amount of the second precursor methylammonium bromide. Additionally, insights are given for the suggested formation mechanism of these anisotropic structures as well as for stability against moisture at ambient conditions in comparison with differently synthesized nanosheet samples.

## INTRODUCTION

Colloidal semiconducting nanocrystals are promising materials for both fundamental research and technical applications, due to their simple chemical synthesis in solution and strong size and shape dependent properties.[1–4] 0D spherical nanoparticles have been investigated extensively in the two decades beginning from 1990 and by modification of the liquid-phase protocols the first simple anisotropic shapes like rods[3], tetrapods[5] or star-like[6] structures were obtained. Nanorods, for example can be achieved in materials with hexagonal symmetry by selective binding of strong surface ligands with a high affinity to the lateral facets and by doing so providing an enhanced reactivity in $c$ direction.[7] Cubic materials are no exception to this and can be forced to generate symmetrical structures like stars, branched particles or cubes.[8–11][12] However, while anisotropic growth resulting in complex structures like wires or sheets is feasible in materials with suitable crystal structures like the hexagonal lattice the cubic structure is limited in regards to simple ligand driven attachment or growth of the seeds via preferred facets. Still, many studies reported the preparation of 2D-like particles based on cubic crystal structures.[13,14] A common synthetic route in colloidal approaches is the template mediated assembly and crystallization of cubic systems.[15–17] In these formation theories a specific combination of parameters has to be set in order to create necessary nano- to microscale reaction domains in the mixture. The templates usually consist of one reactant party, for example metal-cations, which are spatially assembled and confined as lamellar structures by the difference in the polarity of the ligands and solvents used.[16] Adjusting the parameters can provide access to 2D-like structures with a variety in thickness and overall size.[18] Another, less common method to prepare anisotropic structures is a kinetically over-driven reaction in which the high reactivity of the reaction mixture can lead to anisotropic growth.[3,19]

Here, we report about a new preparation method for quasi-2D methylammonium lead bromide nanostripes with a Ruddleson−Popper-type formula $L_2A_{n-1}B_nX_{3n+1}$, where L is a long-chain alkyl and n is the number of neighboring BX monolayers between the organic spacers. To the best of our knowledge, it is the first report of this shape for this material. The dimensions like width, thickness and overall morphology can be tuned with one parameter, respectively. We suggest a template mediated formation mechanism for this cubic material in combination with a kinetic driven reactivity towards the nanostripe shape. Additionally, stability tests were performed for the nanostripes regarding moisture and were compared with nanosheets samples synthesized at different conditions.

**RESULTS AND DISCUSSION**

The MAPbBr$_3$ nanostripes were synthesised by a modified hot injection method reported previously by our group.[20] This method comprises two steps: the preparation of the lead bromide precursor[21] and the formation of the perovskite materials. The precursors already have 2D shape with a distinct ligand shell and are getting dissolved into small nanoparticles and ions. This is followed by a hot injection of the methylammonium bromide (MAB) precursor and the formation of the perovskite structures with quasi-2D morphology. The preparation of the nanostripes is more challenging compared to nanosheets since they have an additional restriction, one in height for the sheet shape and one in the lateral dimensions to form quasi-2D nanostripes. Therefore, the accessibility for this morphology is possible only for a small parameter window. The ideal conditions for the stripe morphology are a reaction temperature of 130 °C, 10 min reaction time, a low amount of main ligand and a small volume of the solvent diphenyl ether (DPE). Variation of each parameter alters the shape of the nanstripes either slightly in height and width or in favour of the formation of nanosheets or 3D structures. One important condition for the formation of the nanostripe structure is the high reactivity of the reaction mixture which was first and foremost achieved by performing the hot-injection step at 130 °C. In our previous report the reactivity was reduced among others by either a lower temperature for the hot-injection step or strongly elevated temperatures among which the initially formed particles dissolved. The formation took place during the slow increase/decrease of the temperature and yielded in nanosheet structures without any side products. In this report, the reactivity is high enough so that the formation of the nanostructures takes place in the moment of the injection. Aliquots taken in the first seconds after the injection reveal an initial formation of about 500 nm long stick-like structures which are not in electronic confinement (Figure S1). The next aliquots taken every 10 seconds present the formation of several micrometers long and a few nanometers thin nanostripes. At the end the initially formed stick-like structures make up for a relatively small part of the product. These results indicate that due to the high reactivity a certain amount of the second precursor reacts quickly after injection and leads to the formation of the stick-like structures. The rest of the precursor is distributed by stirring followed by crystallization of the main product.

The formation mechanism of 2D nanostructures has been in discussion for a long time with approaches like ligand-mediated growth, oriented attachment, or template-driven formation.[22–24] In case of hexagonal crystal structure all mentioned approaches can be present since the facets with different surface reactivities can favor the growth in a distinct direction. The cubic crystal structure on the other hand is highly symmetric in terms of facet distribution and simple ligand-driven synthesis including oriented attachment is unlikely.[16]

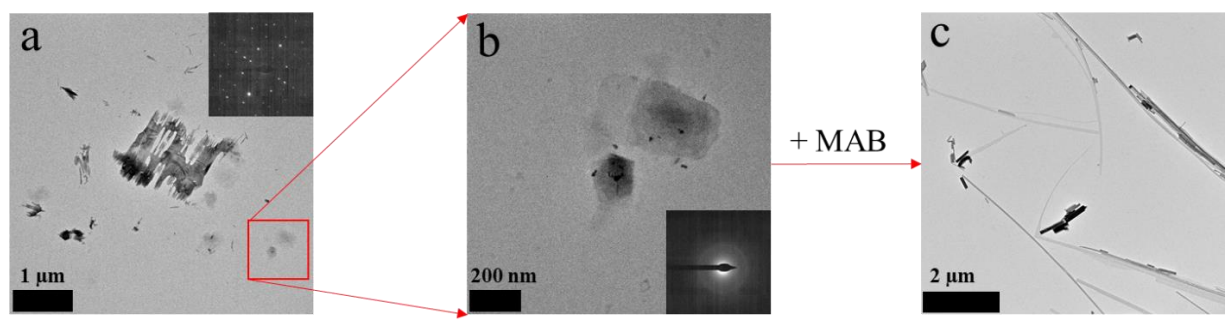

**Figure 1. Observation of the formation process of MAPbBr$_3$ nanostripes. (a, b) Bright-field transmission electron microscopy images of partly dissolved PbBr$_2$ nanostructures surrounded by freshly formed templates for the stripe synthesis prior to the injection of MAB. Insets show SAED images for the corresponding structures. (c) Several micrometer long nanostripes as product in the synthesis 10 min after the injection of MAB.**

Figures 1a-c show two steps in the formation of MAPbBr$_3$ nanostripes evidenced by bright-field transmission electron microscopy (BF-TEM) images. Figures 1a, b depict residue of undissolved PbBr$_2$ structures and potentially freshly formed templates prior to the MAB injection. The inset in Figure 1a displays selected area electron diffraction (SAED) of the PbBr$_2$ with a dot pattern which matches with the reference reflexes of PbBr$_2$.[25] The SAED in Figure 1b shows no reflexes at all which can be explained by a small thickness of these templates or by the absence of crystallinity. After the injection of the MAB precursor and a reaction time of 10 min several micrometers long nanostripes can be obtained.

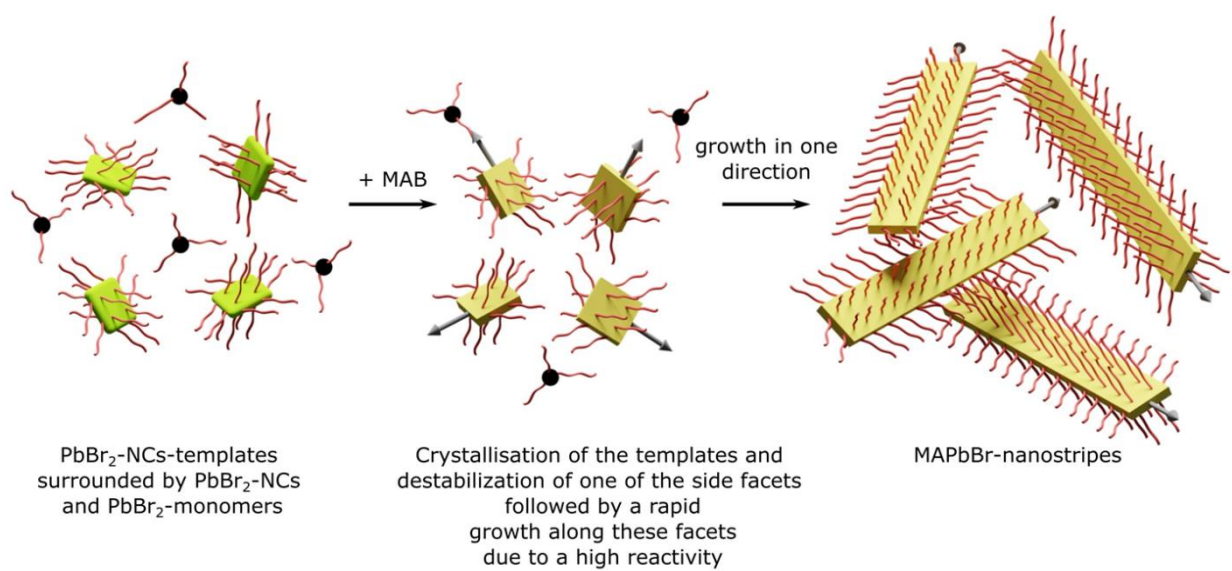

**Scheme 1. Proposed formation mechanism of the MAPbBr$_3$ nanostripe structures.**

Based on these evidences, we suggest a template mediated formation mechanism for the nanostripe structure of cubic MAPbBr$_3$ (Scheme 1). Prior to the injection of MAB the reaction mixture is comprised of rectangular lamellar templates of PbBr$_2$-NCs surrounded by PbBr$_2$-NCs and monomers (Scheme 1 - left). After the addition of MAB, the precursors in the template start to crystallize (Scheme 1 - center). At the same time one of the lateral facets gets destabilized followed by a rapid growth along this facet. Since the reactivity is high and the formation of the nanostripes takes place in the first tens of seconds the growth is performed along the direction of the first destabilized facet (Scheme 1 - right).

Tests regarding reactivity were performed varying the amount of the solvent DPE and the temperature. Changing the amount of DPE means increasing or decreasing the concentration of the reactants and ligands (Figure S2). A higher reactivity (2 mL, 4 mL of solvent) compared to the standard synthesis (6 mL) resulted in samples containing nanostripes with a large variety in width and thickness. A lower reactivity (high amount of DPE, > 6 mL) yielded nanostripes with non-uniform shape along the length and ultimately in perovskite structures with amorphous shapes. A reduced volume with the same amount of reactants and ligands means that the templates are not uniform in size and shape, part of it remains small to yield nanostripes, part of it merges together to form nanostripes with different widths and thicknesses. An increased volume with the same amount of reactants and ligands means that the ligand passivation of the individual templates is not dense enough and growth can take place in more than one direction. The temperature has a huge impact on the formation and stability of the template structures (Figure S3). Lower temperatures around 100 °C or below are not sufficient enough to form stable templates at the given combination of ligands and reactants. As a results, structures with undefined morphology are obtained. Increasing the temperature leads to the appearance of a larger amount of nanostripes. Temperatures higher than that of the standard synthesis (130 °C) yield samples with brick-like structures, stripes and sheets as products. Here, the elevated reactivity leads to unstable templates which gives rise to a large variety in shape and size. In general, we found that the size and stability of the templates is determined by the amount and complex interplay of ligands, reactants and the temperature.

**Impact of DDA concentration.** Dodecylamine (DDA) is a molecule with a saturated 12 carbon chain length and an amine group at one end of the chain which is a typical ligand in the preparation of perovskite nanostructures and plays the role of a L-type ligand.[26] Since it is the main ligand, changes in the amount (between 0.02 mL and 0.6 mL) should alter the size of the nanostripes and in total the morphology of the perovskite nanostructures. Figure 2a depicts the evolution of the morphology of MAPbBr$_3$ nanostructures with the amount of DDA while all other parameters are fixed. A low amount of the amine (0.02 mL) yields 3D brick-like (NBRs) structures which tend to agglomerate. An increase of the amine (0.04 mL) leads to the appearance of nanostripes (NSTRs) as a side product. Further increase results in nanostripes as the sole component. Nanosheet (NS) formation is observable at even higher amounts of the main ligand which ultimately results in products with nanosheets as the sole structures.

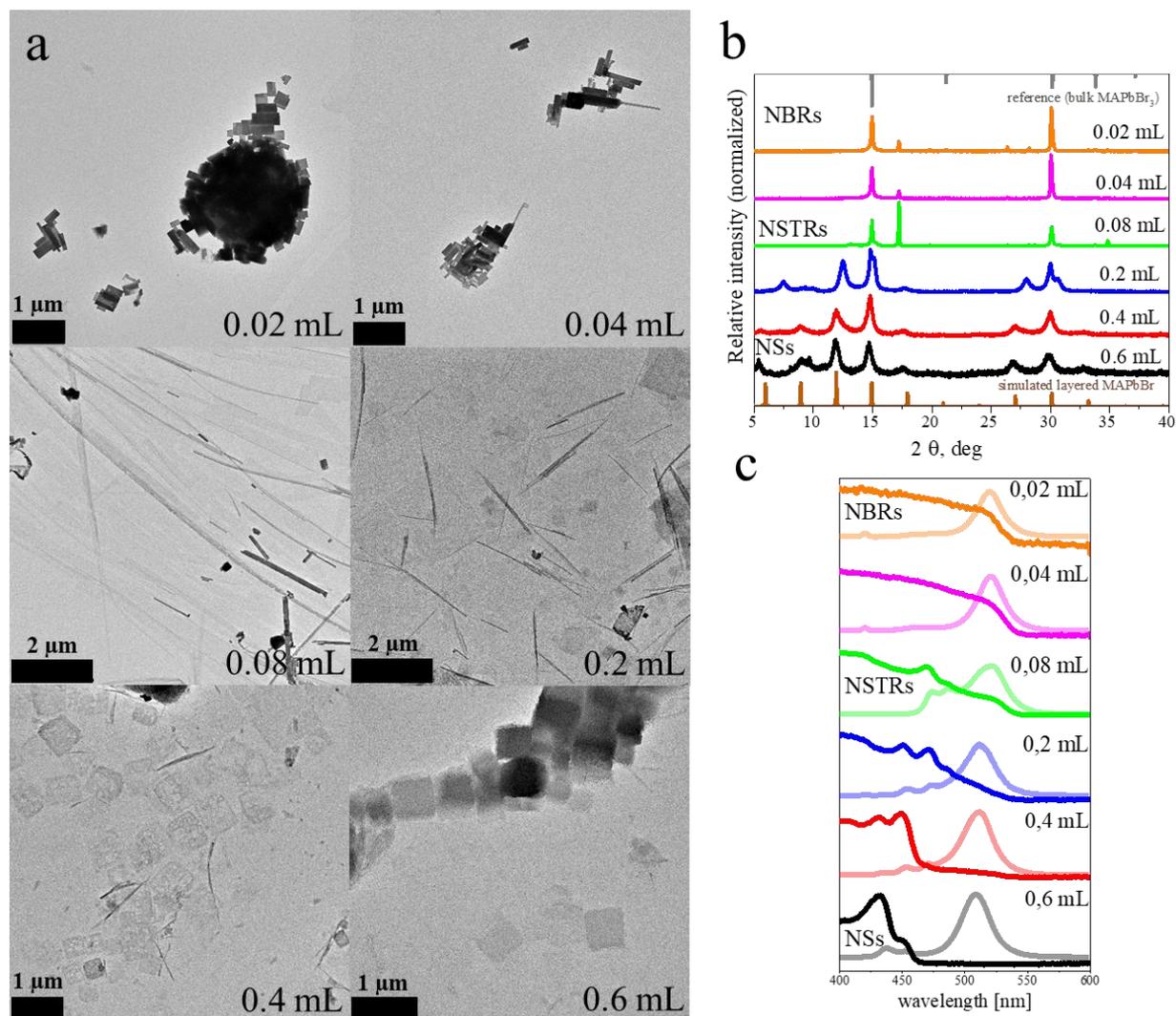

**Figure 2.** Shape variation of MAPbBr$_3$ nanoparticles as a function of amine ligand concentration. The amine used was DDA with a 12-carbon chain length. All other parameters were fixed, and the reaction temperature was 130 °C. The numbers in the images/diagrams correspond to the volume of the DDA-precursor (0.5 g DDA + 4 mL DPE). The reaction time was 10 minutes. (a) BF-TEM images depict the shape evolution of MAPbBr nanoparticles from bricks to stripes to sheets. (b) XRD patterns show a change from bulk to layered structures. (c) UV-Vis absorption (solid) and emission (transparent) spectra reveal the appearance of energy funnelling in the nanosheet structures.[20]

The low amount of DDA meaning less ligands are associated with the precursors and templates which results in an even higher reactivity of the system and leads to the formation of quasi-3D structures. The increase in DDA concentration increases the stability of the precursors and leads to a two-phase reaction in which at first the brick-like structures are formed and afterwards nanostripes. Further increase (0.08 mL) leads to the ideal conditions for the formation of nanostripes. An even higher amount of DDA (0.6 mL) creates stable templates in which the growth of the nanostructures is uniform und yield nanosheets as products. From this point on no stable nanocrystals can be obtained while using higher concentrations of DDA. Here, the initially formed nanostructures are getting dissolved since the high amount of ligands creates strong complexes with the $Pb^{2+}$ and $Br^-$ ions which are highly stable in solution. A change from bulk to layered structures can be observed in the X-ray powder diagrams (XRD) depicted in Figure 2b. The first three samples show an untypical arrangement of the XRD pattern with two unusual signals at 17.1° and 34.8°. The reflexes at 2θ≈15° and 30° (corresponding to (100) and (200) crystallographic planes) and absence of other typical reflexes reveals a strong texture effect and thus supports the statement about the 2D morphology drawn from TEM imaging. In order to understand the origin of the two additional signals in XRD, we performed a post-synthetic treatment, like one additional washing step or the addition of DDA for one of these samples and performed XRD measurements in a capillary (Figure S4). All these measures led to a decrease or disappearance of these signals in intensity compared to the other signals. The XRD pattern of crystalline dodecylamine contains signals at 17 ° and 34°. These signals are part of a set of signals describing a layered structure with periodicity of 31.5 Å observed in the measurement of polycrystalline DDA (see Figure S5). Therefore, in an ideal case all of these signals should be observed in our pattern. However, we observe just two signals supposedly because of cancellation of lower order signals due to stacking containing perovskites structures. Another option could be a situation, that the nanostripes might form stacks in which the dodecylamine crystallizes and is placed oriented owing to the orientation of perovskite NCs on a flat substrate. Such crystallization with a≈5.7 Å, b≈7.1 Å, c=17.7 Å was reported in similar DDA-hydrochloride crystals.[27] The textured XRD would then show only reflexes at 2θ=17 ° and 34 °. In general, we show that the addition of DDA ligand solution to the sample or an additional washing step prior to XRD measurement fully changes the stacking behavior and results in different XRD intensity distribution (Figure S4 b). All the nanosheets which are formed at a higher amount of DDA prefer to organize themselves as stacks. These stacks are highly ordered layered structures consisting of the individual nanosheets. Figure 2c depicts the UV-Vis and PL-spectroscopy data of the corresponding samples and reveals a change from bulk to highly confined structures with a blue shift in the absorption spectra which show a pronounced energy funnelling effect.[20] It has been reported, that the initially formed charge carriers by light excitation migrate to regions with smaller bandgap and recombine. This in turn results in a large red shift from the main absorption to the emission peaks.[28]

To estimate the influence of the amine ligand further we performed the standard synthesis with different amines. All the products yielded 2D structures while the sample obtained with tetradecylamine (TDA) contained the highest amount of nanostripes (Figure S6). Since the chain length is just 2 carbon atoms longer the shape of the template and reactivity was similar to the one with dodecylamine. The biggest difference was observed in the sample prepared with oleylamine where the dominant shape was represented by brick-like structures. In conclusion, it is possible to alter the morphology of MAPbBr$_3$ perovskite materials from brick-like structures to stripes to nanosheets with the variation of just one parameter – the DDA concentration.

**Impact of TOP concentration**

The preparation of a given structure in an anisotropic shape is a complex task on its own which often leaves no room for the manipulation of the dimensionality of these structures like length, height, or width. We found that the co-ligand trioctylphisphine (TOP) can be used as an additional fine-tuning instrument which can alter the width of the nanostripes between about 4 nm to 200 nm. Figure 3 shows the effect of different concentrations of TOP on the morphology. Syntheses without TOP yield products with mixed shapes like nanostripes, shapeless nanocrystals and sheets (Figure 3a). To find out whether all three differently shaped structures are formed at the same time we performed the same synthesis but stopped it right after the injection of the second precursor. Figure S7a reveals the presence of nanosheets and nanostripes. We assume that the shapeless nanocrystals are formed in a second nucleation phase in which the amount of ligands was not sufficient to establish stable templates. Since TOP is a ligand which binds to Pb$^{2+}$ ions the omission of it has an impact on the templates and yields a product with nanosheets as the main shape. Introducing TOP to the synthesis has the effect that nearly no shapeless nanocrystals are formed and that more nanostripes are present (Figure 3b). The templates containing TOP on the surface are stronger separated from the DPE due to a stronger steric nature of this molecule. This explains why increasing the amount of TOP leads to samples which contain solely nanostripes as a product.

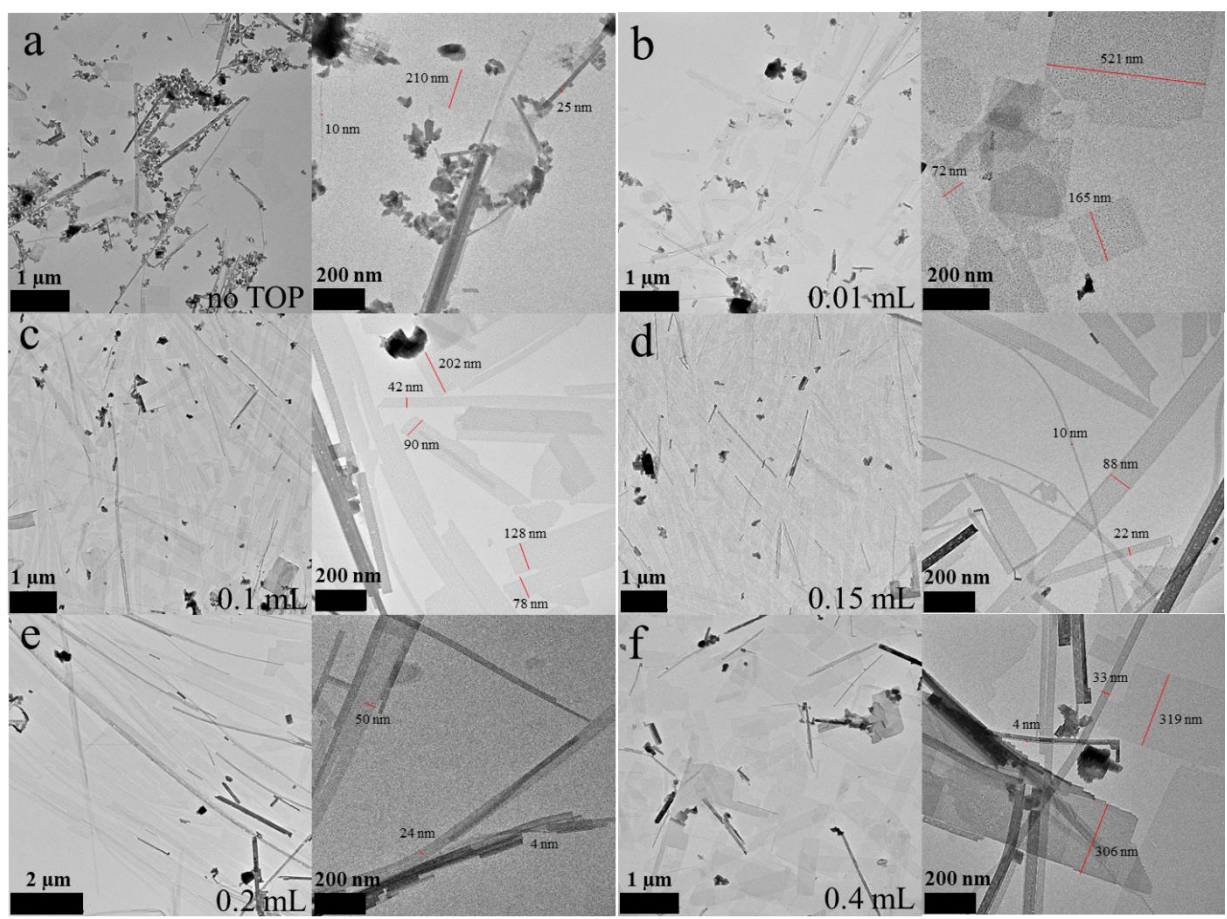

**Figure 3.** Influence of the co-ligand TOP on the morphology in the nanostripe synthesis. The amount of TOP was varied between (a) no TOP, (b) 0.01 mL, (c) 0.1 mL, (d) 0.15 mL, (e) 0.2 mL and (f) 0.4 mL. Left images: overview, right images: zoom in.

A very high amount of TOP means that the reaction mixture is saturated with this molecule and no strong separation of the templates from the rest of the solution can be achieved. In this case most of the product is comprised of MAPbBr$_3$ nanosheets. Figures S8a, b depict XRD, UV-Vis and PL data of these products. The X-ray diagrams show the same signals with the same intensity distribution and a non-layered behavior compared to each other. The only exception is the sample with a high amount of TOP which exhibit one strong additional signal at 12°, which we ascribe to a different kind of stacking of the individual nanocrystals. The UV-Vis and PL spectra show a bulk behavior of the sample without TOP. With the addition of TOP some thinner particles are formed simultaneously. This trend goes on until the sample with a high amount of TOP. Here, just two different thick structures were obtained, which correspond to the nanosheets and nanostripes. TOP is not essential for the formation of the stripe structure but provides good assistance in stabilizing the templates and preventing the appearance of nanocrystals. Most importantly, it can be used to manipulate the width of the nanostripe structures in a range from few to hundreds of nanometers.

**Influence of MAB concentration**

Methylammonium bromide (MAB) is a short organic halide which is often used as the organic part in the perovskite syntheses. In our reaction protocols it is used as the reactant which is added last (after template formation) and starts the reaction. Figure 4 shows TEM and AFM images of MAPbBr$_3$ perovskites prepared with different amounts of the MAB precursor. It is revealed that by varying the amount of MAB the thickness of the perovskite nanostripes can be tuned between about 3 nm to 60 nm. This is highlighted by the contrast in the TEM images and thickness measurements in AFM. A small amount of the MAB precursor yields products with nanosheets, nanostripes and brick-like nanocrystals (NCs). Here, the reactivity is low and only a small part of the templates forms nanostripes, the rest is gradually forming nanosheets. In a second nucleation phase all the residual reactants react to brick-like NCs. Figure S9a depicts a product obtained in the same way but stopped immediately after the injection of the MAB precursor. This sample shows mainly thin anisotropic nanostructures – nanostripes and a small amount of nanosheets. Increasing the amount of the MAB precursor means a higher reactivity which leads to mainly nanostripes as products. Further increase in the amount of the MAB precursor has the effect that the initially formed nanostripes grow in thickness. At first, all the possible templates crystallize and grow in length to form nanostripes and since the reaction mixture has still a sufficient amount of both reactants growth takes place in the out of plane direction. Figures S9 b and c show a product which was prepared in the same way as the 0.08 mL sample but stopped immediately after the injection of the MAB precursor. These nanostripes show much smaller thickness (5 nm to 35 nm) compared to the 10 min version (10 nm to 55 nm). An even higher amount of the MAB precursor had no further influence on both the morphology and thickness. The XRD data shows similar signal distributions for all the samples except for the 0.04 mL sample which depicts a more layered stacking (Figure S7c). This might be due to the higher amount of nanosheets in the product. The UV-Vis and PL data presented in Figure S8d reveal that with a higher amount of MAB precursor a change from confined to bulk structures occurs.

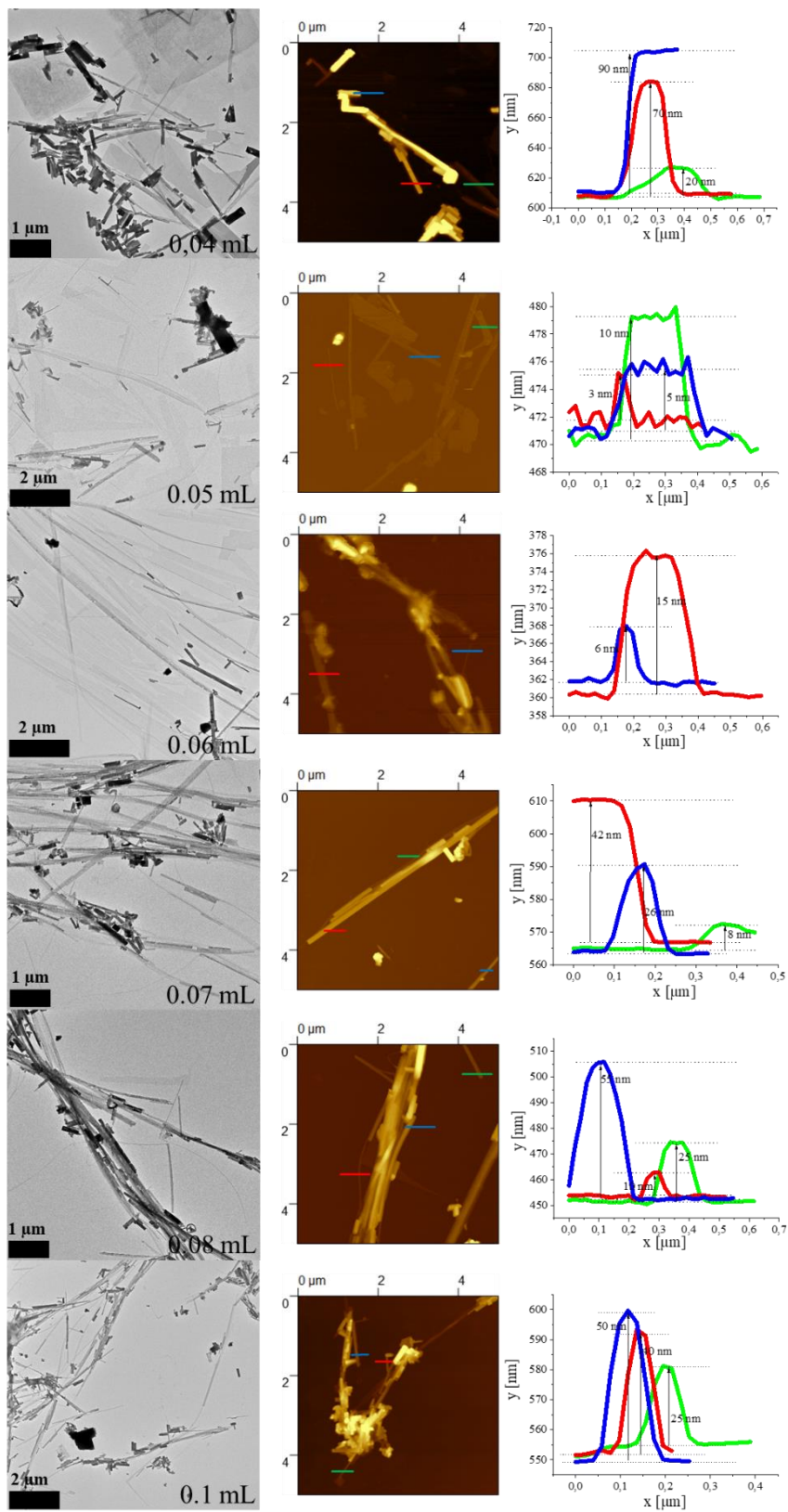

**Figure 4. Shape tuning of MAPbBr₃ nanoparticles as a function of MAB concentration.** The MAB was dissolved in DMF and stored in a glovebox. The numbers in the images correspond to the volume of the MAB precursor. Bright-field TEM images on the left side depict mainly the change in the contrast of the stripe nanostructure with increasing MAB concentration. The reaction time was 10 minutes. The middle and left column show AFM images and the corresponding height profiles of marked nanostripes, confirming the increase in thickness.

**Stability as a function of morphology**

One of the major drawbacks of lead-based perovskite structures is their poor stability against moisture.[29] Thus, we investigated three samples with different morphology or preparation route in terms of stability against moisture over 60 days with UV-Vis spectroscopy and 30 days with XRD at ambient environment. Investigations with longer time frames showed no change in the peak/signal distribution, respectively. The three samples were nanostripes, nanosheets with a minor fraction of nanostripes prepared in the same way as the stripes but with a higher amount of DDA-precursor and nanosheets obtained with an alternative synthesis protocol used as reference.[20] All three samples had similar long average PL lifetimes of 51 ns, 70 ns and 58 ns (Figure S10) as well as similar PL quantum yields of 12 %, 12 % and 8.5 %. The UV-Vis measurements were performed in a way that for every measurement a fresh cuvette was prepared from the stock solution. The XRD data were obtained by measuring the same once drop-casted amount of the material on a silicon wafer. Figure 5 a-c present the UV-Vis results for all three samples.

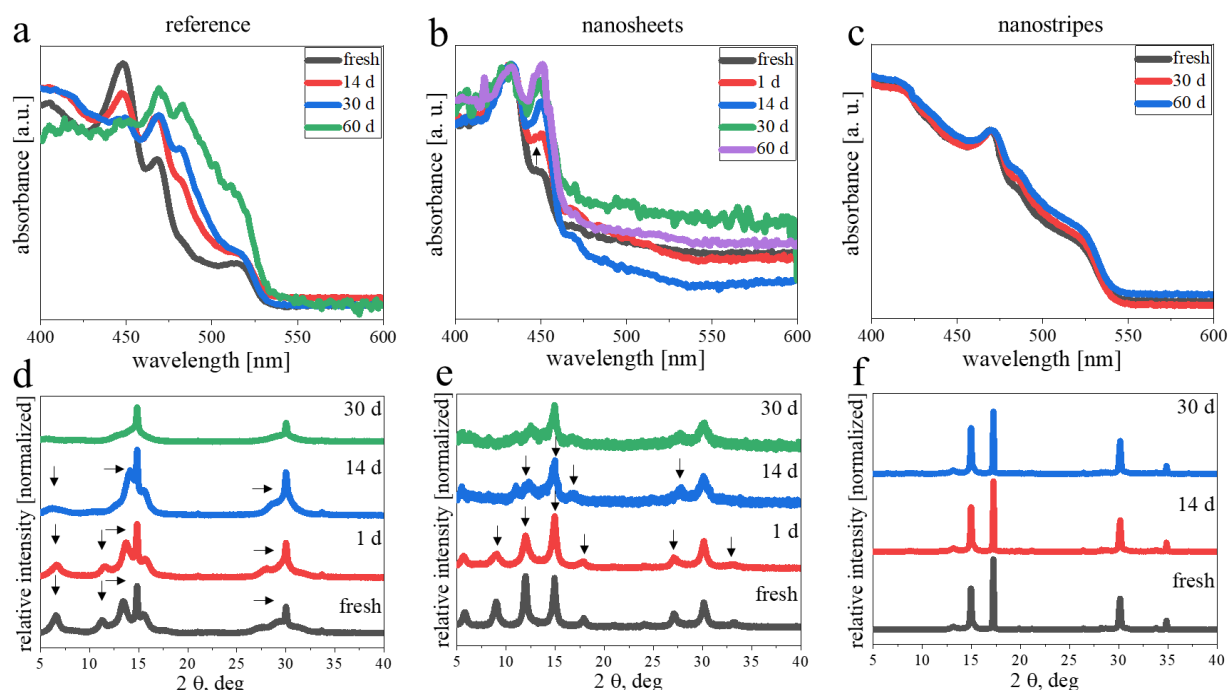

**Figure 5. Stability tests regarding moisture for three different samples of MAPbBr$_3$.** (a-c) UV-Vis absorption spectra of reference nanosheets, nanosheets prepared with the same parameters as the standard nanostripe synthesis but with a higher amount of DDA-precursor and nanostripes. The measurements were performed in a solution state as a function of storage time over 60 days in ambient conditions. (d-f) XRD profiles of the same drop-casted material on a silicon wafer from a fresh state up to 30 days.

The nanosheets obtained with an alternative protocol exhibit a constant change in the whole visible region of the absorption spectrum (Figure 5a). The peak at 450 nm decreases in intensity while all other signals increase. We speculate that the thinner nanosheets are either dissolved and partly serve as additional material for a growth in thickness or nanosheets with different thickness merged and formed thicker structures. The nanosheets prepared in a similar way like the nanostripes show an increase of the peak at 450 nm and a slight decrease of the signal at 430 nm. Here, the same re-crystallization process takes place like for the other nanosheet sample. On the other hand, nearly no change in intensity can be observed for the nanostripe sample in Figure 5c. Figure S11 presents TEM images of the two nanosheets samples one day after the synthesis and two months later. The XRD data depicted in Figure 5 d-f support the evidence obtained by the absorption measurements. With time all signals except of two (14.9° (100) and 30.1° (200)) disappear in the sample with the nanosheets which represents the transition to bulk structures. The nanosheets prepared in a similar way like the nanostripes change from a layered stacking to a non-layered single crystal form. In contrast, the signal and intensity distribution in the nanostripe sample remains unchanged over 30 days.

Based on our observations we can discuss two possible reasons for the higher stability of the nanostripes compared to the nanosheets. One, it could be a different growth direction which results in different facets on the surface. Two, it could be a different ligand passivation on the surface of the particles. To investigate whether the nanostripes exhibit different facets we performed single area electron diffraction (SAED) on single nanostripes and nanosheets (Figure S12). The images show identical patterns for all three samples mentioned above. FTIR spectra of all three mentioned products as well as the used ligands are shown in Figure S13. It is revealed that the nanostripe sample has a stronger passivation with nonanoic acid/nonanoate corresponding to the region between 1150 to 1320 cm$^{-1}$ compared to the nanosheet samples as well as a small shift of the amine group fingerprint at 3000 to 3250 cm$^{-1}$. The signals at 1150 to 1320 cm$^{-1}$ of nonanoic acid/nonanoate are representing the stretching vibration of the C-OH group as well as the wagging vibration of the CH$_2$ group.[30,31] The nanosheet reaction mixture prepared with a similar set of parameters like the nanostripe mixture contains a five times higher amount of amine inside which results in a better passivation by DDA compared to nonanoic acid/nonanoate. The nanosheet reaction mixture prepared with an alternative synthesis protocol contains a 3.4 times higher amount of amine but also 1.9 times more lead bromide precursor which is the source for nonanoic acid/nonanoate compared to the nanostripe mixture. Here, we can observe a more pronounced passivation with the nonanoic acid/nonanoate but still less than for the nanostripe sample. Tawil *et al.* reported about a better stability of cesium lead iodide perovskite nanoparticles in the presence of CTAB ligands.[32] Kim *et al.* discussed the influence of *tert*-butyl alcohol (*t*-BuOH), a short chain molecule with an alcohol group which is present in the antisolvent during the injection of the perovskite precursor.[33] They reported that *t*-BuOH stabilizes the CsPbI$_2$Br structures by passivating defect sites on the surface and therefore increase the moisture stability at ambient conditions. We assume that the stronger passivation of the surface of the nanostripes with the shorter nonanoic acid/nonanoate compared with DDA results in a denser ligand shell which better saturates the dangling bonds and surface defects of the structure. This specific combination of ligands on the surface may only be accessible for the nanostripe shape since the reaction parameter window is limited. Based on our observations, arguments derived from FTIR measurements and plethora of literature we conclude that the ligand passivation is the underlying reason for the discussed differences in stability in our different shaped nanostructures.

**CONCLUSION**

In summary, the formation of MAPbBr$_3$ nanostripes was demonstrated which required a high reactivity of the reaction mixture and was adjusted by performing the synthesis at elevated temperatures, a small reaction volume and a well-balanced ratio and concentration of ligands and reactants. We assign the formation of nanostripes to a template-driven mechanism for all these nanostructures and discuss how reactivity mediated by temperature and concentration can lead to anisotropic growth. Though the parameter window for the formation of the nanostripe shape is narrow it is revealed that by the variation of the amount of TOP and MAB-precursor the size in terms of width and thickness can be tuned, respectively. Additionally, the DDA concentration can be used to give access to 3D brick-like structures, quasi-2D nanostripe particles as well as nanosheets. These findings and the discussion give insight into reactivity and formation behaviour of this material. The nanostripes were compared in terms of stability against moisture with two differently synthesized nanosheets samples. The nanostripes demonstrated a superior stability monitored by UV-Vis and XRD measurements while exhibiting similar optical properties in terms of PL lifetime and PLQY. We attribute this behavior to a stronger ligand surface passivation by nonanoic acid.

**EXPERIMENTAL SECTION**

**Chemicals and reagents.** All chemicals were used as received: Lead(II) acetate tri-hydrate (Aldrich, 99.999%), nonanoic acid (Alfa Aesar, 97%), tri-octylphosphine (TOP; ABCR, 97%), 1-bromotetradecane (BTD; Aldrich, 97%), oleylamine (ACROS, 80−90%), methylammonium bromide (MAB; Aldrich, 98%), diphenyl ether (DPE; Aldrich, 99%), toluene (VWR, 99,5%), dimethylformamide (DMF; Aldrich, 99,8%), octadecylamine (ODA; Aldrich, 97%), octylamine (OA; Aldrich, 99%), hexadecylamine (HDA; Aldrich, 90%), dodecylamine (DDA; Merck, 98%), tetradecylamine (TDA; Merk, 98%).

**$PbBr_2$ nanosheet synthesis.** In a typical synthesis a three neck 50 mL flask was used with a condenser, septum and thermocouple. 860 mg of lead acetate tri-hydrate (2.3 mmol) were dissolved in 10 mL of nonanoic acid (57 mmol) and 10 mL of 1-bromotetradecane (34 mmol) and heated to 75 °C until the solution turned clear in a nitrogen atmosphere. Then vacuum was applied to remove the acetic acid which is generated by the reaction of nonanoic acid with the acetate from the lead precursor. After 1.5 h the reaction apparatus was filled with nitrogen again and the reaction was started by adding 0.06 mL of TOP (0.13 mmol) at a temperature of 140 °C and was stopped 8 minutes later. After 8 minutes the heat source was removed and the solution was left to cool down below 80 °C. Afterwards, it was centrifuged one time at 4000 rpm for 3 minutes. The particles were suspended in 7.5 mL toluene and put into a freezer for storage.

**Synthesis of $MAPbBr_3$ nanostripes.** A three neck 50 mL flask was used with a condenser, septum and thermocouple. 6 mL of diphenyl ether (37.7 mmol), 0.08 mL of a 500 mg dodecylamine (2.69 mmol) in 4 mL diphenyl ether precursor were heated to 80 °C in a nitrogen atmosphere. At 80 °C 0.2 mL of trinoctylphosphine (0.45 mmol) was added. Then vacuum was applied to dry the solution. After 1.5 h the reaction apparatus was filled with nitrogen again and the temperature was set to 130 °C. 0.8 mL of as prepared $PbBr_2$ nanosheets in toluene were added and heated until everything was dissolved. The synthesis was started with the injection of 0.06 mL of a 300 mg methylammonium bromide (2.68 mmol) in 6 mL dimethylformamide precursor. After 10 minutes the heat source was removed and the solution was left to cool down below 60 °C. Afterwards, it was centrifuged at 4000 rpm for 3 minutes. The particles were washed two times in toluene before the product was finally suspended in toluene again.

**Synthesis of $MAPbBr_3$ nanosheets.** A three neck 50 mL flask was used with a condenser, septum and thermocouple. 10.5 mL of diphenyl ether (66.1 mmol), 0.06 mL of oleylamine (0.18 mmol) were heated to 80 °C in a nitrogen atmosphere. At 80 °C 0.25 mL of trinoctylphosphine (0.56 mmol) was added. Then vacuum was applied to dry the solution. After 1.5 h the reaction apparatus was filled with nitrogen again and the temperature was set to 120 °C. 1.5 mL of as prepared $PbBr_2$ nanosheets in toluene were added and heated until everything was dissolved. The temperature was reduced to 35 °C and the reaction was started with the injection of 0.06 mL of a 300 mg methylammonium bromide (2.68 mmol) in 6 mL dimethylformamide precursor. The temperature was slowly increased over a period of 20 minutes to reach 160 °C. After this 20 minutes the heat source was removed and the solution was left to cool down below 60 °C. Afterwards, it was centrifuged at 4000 rpm for 3 minutes. The particles were washed two times in toluene before the product was finally suspended in toluene again.

**Characterization.** The TEM samples were prepared by diluting the nanostripe suspension with toluene followed by drop casting 10 µL of the suspension on a TEM copper grid coated with a carbon film. Standard images were done on a Talos-L120C and EM-912 Omega with a thermal emitter operated at an acceleration voltage of 120 kV and 100 kV. X-ray diffraction (XRD) measurements were performed on a Panalytical Aeris System with a Bragg-Brentano geometry and a copper anode with a X-ray wavelength of 0.154 nm from the Cu-kα1 line. The samples were measured by drop-casting the suspended nanostripes on a <911> or <711> cut silicon substrate. Atomic force microscopy (AFM) measurements were performed with AFM from Park Systems XE-100 in non-contact mode. UV/vis absorption spectra were obtained with a Lambda 1050+ spectrophotometer from Perkin Elmer equipped with an integration-sphere. The PL spectra measurements were obtained by a fluorescence spectrometer (Spectrofluorometer FS5, Edinburgh Instruments). For the time-resolved PL measurements, a picosecond laser with 375 nm excitation wavelength and 100 kHz repetition rate was used. The decay profiles are tail-fitted with a tri-exponential function $R(t) = A_1 \exp\left(-\frac{t}{\tau_1}\right) + A_2 \exp\left(-\frac{t}{\tau_2}\right) + A_3 \exp\left(-\frac{t}{\tau_3}\right)$, and the average PL lifetime is calculated using the formula $\tau_{average} = \frac{A_1\tau_1^2 + A_2\tau_2^2 + A_3\tau_3^2}{A_1\tau_1 + A_2\tau_2 + A_3\tau_3}$. PLQYs of the samples were measured using an absolute method by directly exciting the sample solution and the reference (toluene in our case) in an SC-30 integrating sphere module fitted to a Spectrofluorometer FS5 from Edinburg Instrument. During the measurement, the excitation slit was set to 6.5 nm, and the emission slit was adjusted to obtain a signal level of $1\times10^6$ cps. A wavelength step size of 0.1 nm and an integration time of 0.2 s were used. The calculation of absolute PL QY is based on the formula, $\eta = \frac{E_{sample} - E_{ref}}{S_{ref} - S_{sample}}$, where $\eta$ is absolute PL QY, $E_{sample}$ and $E_{ref}$ are the integrals at the emission region for the sample and the reference, respectively, and $S_{sample}$ and $S_{ref}$ are the integrals at the excitation scatter region for the sample and the reference, respectively. The selection and calculation of integrals from the emission and excitation scattering region and the calculation of absolute PL QY were performed using the FLUORACLE software from the Edinburg Instrument. FTIR measurements were performed by drying the nanomaterials and putting the powders on a diamond-ATR unit (PerkinElmer Lambda 1050+). The FTIR measurements are performed with a range from 450 to 4000 cm$^{-1}$.

## ASSOCIATED CONTENT

**Supporting Information**

Additional TEM images, PL decays and corresponding fits, tables with fitting parameters, FLIM images.

## ACKNOWLEDGMENTS

The authors thank the Sylvia Speller group and Regina Lange for providing the AFM setup. We thank our co-worker Ronja Piehler for designing the Scheme 1. Deutsche Forschungsgemeinschaft (DFG,

German Research Foundation) is acknowledged for funding of SFB 1477 "Light-Matter Interactions at Interfaces", project number 441234705, W03 and W05. We also acknowledge the European Regional Development Fund of the European Union for funding the PL spectrometer (GHS-20-0035/P000376218) and X-ray diffractometer (GHS-20-0036/P000379642) and the Deutsche Forschungsgemeinschaft (DFG) for funding an electron microscope Jeol NeoARM TEM (INST 264/161-1 FUGG) and an electron microscope ThermoFisher Talos L120C (INST 264/188-1 FUGG).